\documentclass[aps,prb,twocolumn,showpacs,10pt]{revtex4-1}
\usepackage[dvips]{graphicx}
\usepackage{mathrsfs}
\usepackage{amsmath}
\usepackage{amssymb}
\usepackage{amsfonts}
\usepackage{ae}
\usepackage{bm}
\usepackage{units}
\usepackage{natbib}

\usepackage{xcolor}
\usepackage{hyperref}
\hypersetup{
    colorlinks,
    linkcolor={blue!75!black!80!yellow},
    citecolor={blue!75!black!80!yellow},
    urlcolor={blue!75!black!80!yellow}
}

\begin{document}
\title{Plasmonics for emerging quantum technologies}

\author{Sergey~I.~Bozhevolnyi$^1$ \& N.~Asger~Mortensen$^{2}$}

\affiliation{$^1$Centre for Nano Optics, University of Southern Denmark, Campusvej 55, DK-5230~Odense~M, Denmark\\
$^2$Department of Photonics Engineering, Technical University of Denmark, DK-2800 Kongens Lyngby, Denmark}

\date{\today}

\maketitle

Expanding the frontiers of information processing technologies and, in particular, computing with ever increasing speed and capacity has long been recognized an important societal challenge, calling for the development of the next generation of quantum technologies\cite{Gibney:2016}. With its potential to exponentially increase computing power, quantum computing opens up possibilities to carry out calculations that ordinary computers could not finish in the lifetime of the Universe, while optical communications based on quantum cryptography become completely secure. At the same time, the emergence of Big Data and the ever increasing demands of miniaturization and energy saving technologies bring about additional fundamental problems and technological challenges to be addressed in scientific disciplines dealing with light-matter interactions. In this context, quantum plasmonics represents one of the most promising and fundamental research directions\cite{Tame:2013,Bozhevolnyi:2016a} and, indeed, the only one that enables ultimate miniaturization of photonic components for quantum optics when being taken to extreme limits in light-matter interactions (Fig. 1). While plasmon phenomena are inherently of quantum nature, surface plasmons (SPs)\cite{Maier:2007} have for decades been successfully explored with a mindset of classical electrodynamics\cite{Engheta:2015} (lower, left corner). However, fundamental problems found in all corners of this schematic chart (Fig.~1) increase exponentially their complexity when coming to experience also fundamental size limitations posed by the very nature of both light and matter, both consisting of quanta and thereby being discrete. With this perspective, we anticipate an emerging new era of \emph{quantum plasmonics} (upper, right corner), where both light and matter exhibits quantum mechanical effects. We expect numerous challenges emerging out of the \emph{treasure trove}, while also providing the fuel for emerging quantum technologies. Below, we discuss some intriguing aspects of the different regimes outline in Fig.~1.

\begin{figure*}
\centering
\includegraphics[scale=0.8]{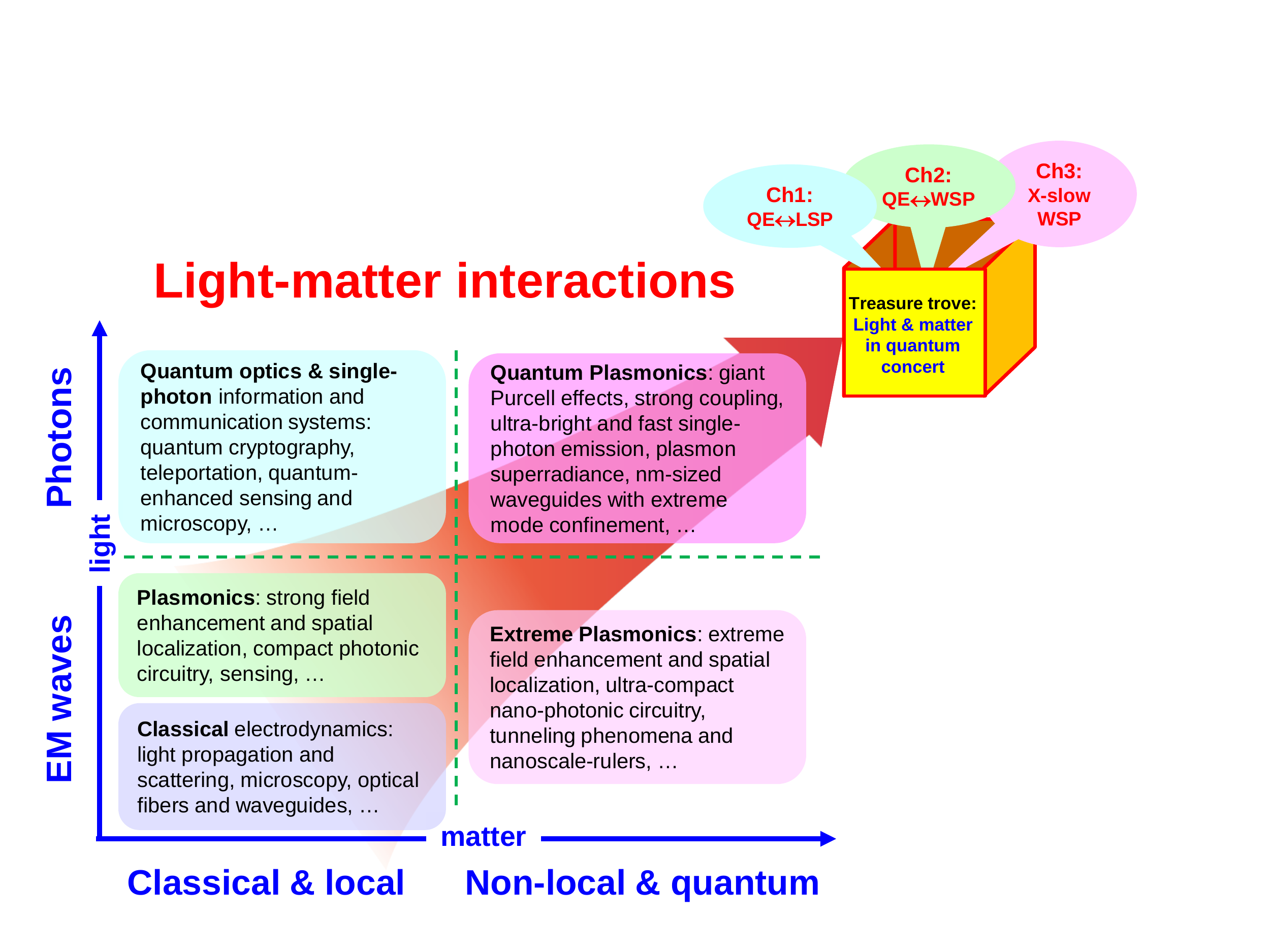}
\caption{Quantum aspects of plasmonics exhibit in both the quantum nonlocal response of matter as well in the quantized light fields. With the developments of quantum plasmonics, we anticipate a \emph{treasure trove} of emerging quantum technologies that harvest from the quantum concert of both light and matter. In the main text we outline 3 challenges associated with this.}
\label{fig1}
\end{figure*}

Recently, there has been significant efforts in what we here term \emph{extreme plasmonics} (lower, right corner) where extremely confined plasmonic modes\cite{Gramotnev:2010} are promoting electrodynamics beyond the accurate description of the local-response approximation of light-matter interactions\cite{Raza:2015a}. Intriguing and classically unexpected experimental observations include frequency blueshifts of plasmon resonances in few-nanometer silver nanoparticles\cite{Scholl:2012,Raza:2015c} and gold plasmonic gap structures\cite{Ciraci:2012} as well as gap-dependent broadening and onset of charge-transfer plasmons in dimers and gap structures with sub-nanometer gaps\cite{Savage:2012,Mertens:2013,Tan:2014}. Theoretical explanations range from hydrodynamic nonlocal response\cite{Mortensen:2014,Toscano:2015} to ab-initio quantum mechanics\cite{Zhu:2016,Varas:2016} accounting for quantum effects such as quantum pressure waves, Landau damping, quantum spill-out and tunneling. In the other extreme, \emph{quantum optics \& single-photon phenomena} (upper, left corner) have been interfaced and enhanced by the classical electrodynamics of plasmonic nanostructures\cite{Akimov:2007,Bermudez-Urena:2015}. Here, the quantumness is in the non-classical states of the electromagnetic field, or in the quantum properties of the light emitters\cite{Chang:2006,Huck:2016}. 

It is the merging of quantum optics with extreme plasmonics that now drives us to new exciting \emph{quantum plasmonics} (upper, right corner), being quantum in a double sense: Both light and matter exhibit quantum mechanical effects! As an appetizer from this almost virgin territory of quantum plasmonics, we here mention the recent exploration of single-molecule strong coupling dynamics in sub-nanometer plasmonic gap cavities\cite{Chikkaraddy:2016}. 

Many crucial issues in modern plasmonics revolve around the fact that the most exciting and unique feature of plasmonic modes, viz., the feasibility of broadband and extreme (down to atomic scale) mode confinement as well as associated enormous field enhancement by use of metal nanostructures, is directly associated with the energy loss (via radiation absorption) in metal that progressively increases with the field confinement\cite{Khurgin:2015}. The fundamental problem is then to understand to what extent this unique feature, enabling both ultimate miniaturization\cite{Gramotnev:2014} and ultra-strong coupling to quantum emitters (QE)\cite{Pelton:2015}, should be exercised before the inevitable energy loss destroys the outcome of its exercise. Enormous challenges emerge already on the way to proper formulation of the associated problems, since the whole well-developed macroscopic treatment of light-matter interactions breaks down and no longer works on the nanoscale, requiring non-local and quantum effects to be taken into account.

The whole set of fundamental problems associated with extreme plasmonics can be factorized into several tradeoffs representing conflicting tendencies that are being enormously enhanced when reaching extreme limits. Careful analysis and assessment of these tradeoffs is indispensable for accurate mapping of the field boundaries and potential developments. In this context, the exploration of fundamental limits in light-matter interactions can be categorized into several major research challenges that are all related to the aforementioned key issues. 

\emph{Challenges.} It is known\cite{Pelton:2015} that placing QEs in optical resonators with high quality factors ($Q$) and/or very small volumes ($V$) results in the spontaneous emission (SE) enhancement described by the Purcell factor $F \sim Q/V$. While extremely high $Q$ values ($> 10^6$) can be attained with photonic crystal micro-cavities, the rate of single-photon QE emission out of such a resonator is inevitably limited due to its high quality factor\cite{Pelton:2015}. Alternatively, localized surface-plasmon (LSP) excitations enable extreme light confinement (down to nanometer-scale), providing thereby unique possibilities for SE enhancement\cite{Tame:2013}. Quality factors of SP-resonators are, contrary to the previous case, rather limited ($Q < 100$), and ultra-bright single-photon sources become feasible by squeezing the LSP volume ($V \rightarrow 0$). At the same time, extreme LSP confinement leads to increasing QE quenching and thus decreasing the QE quantum yield, so that the issue of ultimate SE enhancement remains open (Ch1). It should be noted that the most promising results (ultrafast SE emission) have been obtained with special SP-resonators, viz., metal-insulator-metal nanocavities supporting gap SPs\cite{Hoang:2015}, whose unique features require very careful theoretical analysis with nonlocal and quantum effects being taken into account.

The problem of SE enhancement in the case of QE coupled to propagating waveguide surface plasmons (WSP) is somewhat similar to and yet very much different from the above (Ch2). In this case, the Purcell factor is determined by the ratio between the waveguide mode group index and the mode area, i.e. $F\sim n_g/S_m$. While very high group indexes ($>10^2$) can be attained with photonic crystal waveguides, the bandwidth available for QE emission into such a waveguide is inevitably limited due to its proximity to the (prohibited) band gap. Contrary to that, various WSP modes, including the gap and channel SP modes\cite{Smith:2015}, can be confined (practically at any wavelength) within extremely small cross sections, increasing thereby the QE-WSP coupling efficiency\cite{DeLeon:2012}. At the same time, however, the loss-related problems would also become significant. In fact, one should simultaneously maximize the Purcell factor, the coupling to propagating SP modes (rather than to lossy SPs) and the normalized SP propagation length\cite{Bermudez-Urena:2015}. The record of the corresponding product is currently $\sim 6.6$\cite{Bermudez-Urena:2015}, but its fundamental limit is yet to be established. A novel issue in this context is the problem of unidirectional and efficient SE into WSP modes\cite{Curto:2010}, since one would ideally couple all emitted photons into WSP quanta propagating in the same direction. 

Another unique feature of SP-based waveguides (Ch3) is related to the possibility of seamless interfacing of electronic and photonic circuits by employing the same metal circuitry for both guiding the optical radiation and transmitting the electrical signals that control the guidance\cite{Ebbesen:2008}. The latter implies that metal electrodes used for radiation control are located at the maximum of the SPW mode intensity (reached at the metal-dielectric interface), maximizing the controlling efficiency and thus ensuring considerably lower power requirements. This feature opens unique perspectives for substantial reductions in sizes and energy of SP-based photonic components and circuits while extending the operation bandwidth\cite{Haffner:2015}. It also allows one for the realization of conceptually new functionalities and exploitation of new materials with fascinating properties, such as single-crystaline gold films\cite{Wu:2015} or graphene\cite{Grigorenko:2012}, as e.g. demonstrated by realizing a hybrid SP-graphene waveguide modulators\cite{Ansell:2015}. The degree of WSP modulation can be enhanced by squeezing the WSP mode area and increasing thereby the slowdown factor. The associated increase in the WSP absorption has to be carefully investigated and analyzed so as to take advantage of the enormous potential of the aforementioned unique WSP characteristics. Proper analysis requires further development of the theoretical models tailored suitably for dealing with nanostructured light-matter interactions.

\emph{Spinoff directions.} Investigations of underlying physics and fundamental limitations associated with LSP excitations would be greatly beneficial for the whole field of flat optics based on phase and amplitude-gradient metasurfaces\cite{Kildishev:2013}, allowing one to mould the radiation with an unprecedented control over its polarization and propagation characteristics as well as produce surface coloration at nanoscale\cite{Kristensen:2016}. In the field of sustainable energy sources, LSP-induced resonance energy transfer becomes increasingly important for solar energy conversion\cite{Li:2015}. Moreover, recent ground-breaking discoveries emphasize that even inevitable light absorption in plasmonics can be turned into gain within various topics: from plasmon-enhanced optical tweezing to thermal therapy and data storage\cite{Ndukaife:2016}. Finally, mastering unique plasmonic features associated with the excitation of both LSP and WSP would most certainly provide additional possibilities for pairing photons with phonons in quantum opto-mechanics\cite{Riedinger:2016}. It is thus clear that studies related to the above objectives would have major consequences and implications to many sub-fields of modern nanoscience and emerging quantum technologies\cite{qurope} including radiation-lifetime engineering, plasmon-enhanced chemistry, single-molecule sensing, quantum microscopy, quantum optics and opto-mechanics in general, and single-photon sources, in particular.

\emph{Acknowledgments.} S.~I.~B. acknowledges financial support by European Research Council, Grant 341054 (PLAQNAP), while N.~A.~M. was supported by the Danish Council for Independent Research--Natural Sciences (Project 1323-00087).

\bibliographystyle{apsrev4-1}

\bibliography{references}

\end{document}